\documentclass[12pt]{iopart}
\usepackage{graphicx}
\usepackage{amssymb}
\usepackage{amsfonts}
\usepackage{dcolumn}
\usepackage{epsfig}
\usepackage{subfigure}
\usepackage{bm}
\usepackage{calc}
\usepackage{braket}

\begin{document}

\title{Trapping and detection of single atoms using a spherical mirror}

\author{Arpan Roy, Andrew Bah Shen Jing and Murray D. Barrett}

\address{Centre for Quantum Technologies and Department of
  Physics, National University of Singapore, 3 Science Drive 2, 117543 Singapore.}
\ead{phybmd@nus.edu.sg}
\begin{abstract}

We fabricate a miniature spherical mirror for tightly focusing an optical dipole trap for neutral atoms. The mirror formation process is modelled to predict the dimensions for particular fabrication parameters. We integrate the spherical mirror with a neutral atom experiment to trap and detect a single atom with high efficiency. The mirror serves the dual purpose of focusing the dipole trap as well as collection of the atomic fluorescence into an optical fibre.

\end{abstract}
\newpage
\tableofcontents
\newpage

\maketitle

\section{Introduction}

For quantum information applications based on atoms, large atom-photon interactions can be used to interface photonic and atomic qubits.  Enhancing the interaction can be achieved using either high finesse optical cavities \cite{Hinds2} or strongly focused laser fields \cite{Christian}. High finesse cavities enhance the atom-photon interaction by repeated reflection of a single photon so that the field from a single photon is enhanced by the cavity finesse. Although this approach has been successful with neutral atoms, cavity experiments are complicated to implement for the ion trap system due to technical problems associated with proximity of the cavity mirrors to the ion \cite{Isaac2}. Cavity based systems are also difficult to scale up to a large system of atom-photon interfaces \cite{lepert2011arrays}. In contrast, the strong focusing approach is easy to implement \cite{weber2006analysis} and a number of approaches have been considered towards scalability.

For the purposes of scaling to a large number of atoms it also desirable to provide the means to trap and individually address single atoms.  A large number of individually addressable single atoms have been trapped by spatial light modulators combined with custom made microscope objectives \cite{bergamini2004holographic} but the method is neither simple nor low cost. Micro lens arrays have also been used to create arrays of approximately 80 individually addressable optical dipole traps \cite{dumke2002micro} but ensuring single atom occupancy was not possible.  An alternative approach is to use spherical mirrors.

Spherical mirrors are easy to integrate with both ion traps \cite{Blinov1} and neutral atom experiments.  In the case of neutral atoms, an optical dipole trap can be formed by focusing collimated light with the spherical mirror \cite{Hinds3}. If the radius of curvature of the mirror is small, the focus is tight enough to engage the collisional blockade phenomenon \cite{schlosser2002collisional} which ensures only one atom can be trapped under weak loading conditions. Alternatively, light assisted collisions in a tightly confining dipole trap can be used to prepare single atoms \cite{grunzweig2010near}. Moreover, the interaction light can be focused by the same mirror to a tight spot for free space atom-photon coupling. Although spherical mirrors suffer from aberrations, these can be corrected with proper optical elements outside the vacuum chamber \cite{Blinov1} and become less significant with decreasing mirror dimensions.  For sufficiently small radius of curvatures, there can be a significant coupling of the light reflected by the mirror into a single mode fibre \cite{van2012efficient}.

Current methods of producing spherical mirrors involved either silicon based fabrication \cite{Hinds1} or conventional glass polishing techniques. Although arrays of spherical mirrors can be produced by silicon fabrication method, the radius of curvature is limited to hundreds of microns making it difficult to integrate with any practical atom trapping experiments. Conventional glass polishing produced mirrors of large dimensions and cannot be used to tightly focus trapping light. Recently we developed a fabrication technique that bridges the gap between the existing technologies, producing mirrors with large enough dimensions for easy integration with experiments but small enough to enable tight focusing of the dipole trap \cite{roy2011fabrication}.

In this paper we demonstrate the use of our spherical mirrors for trapping single atoms at the focus of the mirror. We obtain a tightly focusing dipole trap with a waist of $\sim 2\,\mathrm{\mu m}$ by illuminating the mirror with a collimated $850\,\mathrm{nm}$ laser. We use light assisted collisions \cite{grunzweig2010near} to prepare single atoms in this trap with $\sim 70\%$ efficiency. Fluorescence collected using the mirror is coupled into a multi-mode optical fibre giving an atom detection efficiency of $99\%$ in a $100\,\mathrm{\mu s}$ integration time.

The paper is organized as follows. We start with a model of the fabrication process, which illustrates how the mirror dimensions can be reliably controlled and predicted.  We then characterize the focussing capability of the mirror.  Details of our experimental set up are then given in which we characterize the optical dipole trap in terms of single atom loading and detection efficiency.

\begin{figure}[t]
     \centerline{\includegraphics[width=1\textwidth]{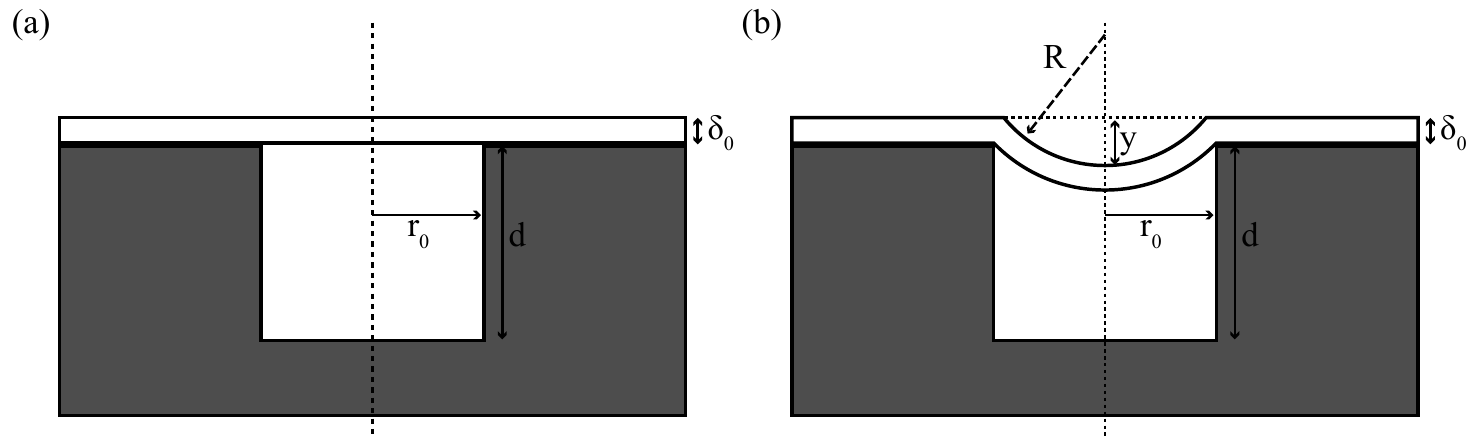}}
         \caption{\label{melting}The fabrication process. (a) Glass slide of thickness $\delta_{0}$ placed on cylindrical hole of radius r$_{0}$ and depth h. (b) After deformation, the dimple has a depth d and a radius of curvature R. In case of the mirror for the optical dipole trap, typical dimensions are $\delta_{0}=150\,\mathrm{\mu m}$ and $r_0 \sim d \sim$ 2 mm.}

\end{figure}

\section{Fabrication and Modelling}

Details of the mirror fabrication process has been reported elsewhere \cite{roy2011fabrication}. Briefly, the method uses borosilicate D263 glass and a ceramic substrate (Macor) \cite{Corning}. Cylindrical holes are drilled into the ceramic and a borosilicate cover slip is placed over them. The substrate and the glass is then heated in a furnace at low pressure.  When heated to $800\,\mathrm{^oC}$, the glass softens and strongly adheres to the substrate sealing the holes at low pressure. Air at higher pressure is then introduced and the pressure difference causes the softened glass to curve inwards creating smooth spherical dimples. The process results in ultra smooth surfaces with sub-nanometer roughness.

To estimate the mirror parameters obtained using this process, we adopt the model developed in \cite{eklund2007glass} for the fabrication of glass micro-spheres. Throughout the deformation process, we assume that the glass surface maintains a near spherical shape. Thus, the radius of curvature, $R$, is given by
\begin{equation}
\label{roc}
R = \frac{y^2 +r_{0}^2}{2y},
\end{equation}
where $r_{0}$ is the radius of the cylindrical hole and $y$ is the depth of dimple as illustrated in Fig.\ref{melting}(b).  To determine $y$, we assume that the entire process takes place at a constant temperature. Thus, the pressure, $P$, and volume, $V$, of the trapped gas can be related to the initial values by the ideal gas law
\begin{equation}
 P V=P_\mathrm{0}V_\mathrm{0}.
\label{gaslaw}
\end{equation}
From geometric considerations the volumes $V$ and $V_0$ are given by
\begin{eqnarray}
\label{vol}
V_\mathrm{0} &=& \pi r_{0}^2 d,\\
V & = & V_{0} - V_\mathrm{d} \nonumber\\
& = & \pi r_{0}^2 d - \pi (\frac{y^3}{6}+\frac{r_{0}^2 y}{2}),
\end{eqnarray}
where $V_\mathrm{d}$ is the volume of the dimple.  Thus, the pressure of the trapped gas is given by
\begin{equation}
\label{pressure}
P = \frac{P_{0} V_{0}}{V} = P_{0} \frac{\pi r_{0}^2 d }{\pi r_{0}^2 d - \pi (\frac{y^3}{6}+\frac{r_{0}^2 y}{2})}
\end{equation}
giving a pressure difference, $\Delta P=P_\mathrm{ext}-P$, of
\begin{equation}
\label{pressurediff}
\Delta P = P_{\mathrm{ext}} - P_{0} \frac{\pi r_{0}^2 d }{\pi r_{0}^2 d - \pi (\frac{y^3}{6}+\frac{r_{0}^2 y}{2})}.
\end{equation}
where $P_\mathrm{ext}$ is the external pressure introduced after the holes were sealed.  Setting $\Delta P = 0$ then provides an equation for the equilibrium depth $y$ as a function of $r_0$, $P_0$ and $P_\mathrm{ext}$.  The solution can then be used in Eq.~\ref{roc} to determine the radius of curvature.

The time variation of $y$ can also be modelled using the methods given in \cite{eklund2007glass}. The softened glass at $800\,\mathrm{^oC}$ behaves like an incompressible Newtonian fluid and it can be shown that time variation of $y$ satisfies
\begin{equation}
\label{rateofdepth}
\frac{\mathrm{d}y}{\mathrm{d}t} = \frac{1}{24 \eta r_{0}^2 \delta_{0}} \frac{(r_{0}^2 + y^2)^3}{y^2} (\Delta P),
\end{equation}
where $\eta$ is the viscosity of borosilicate glass at $800\,\mathrm{^oC}$ and $\delta_{0}$ is the thickness of the glass.  Taking $\eta=10^{6.6}\,\mathrm{Pa.s}$ \cite{Schott}, $\delta_{0}=150\,\mathrm{\mu m}$, $r_0 \sim d \sim y \sim 1\,\mathrm{mm}$ and $\Delta P\sim 150\,\mathrm{mbar}$ gives an estimated time scale of several minutes.

\begin{figure}[t]
     \centerline{\includegraphics[width=0.9\textwidth]{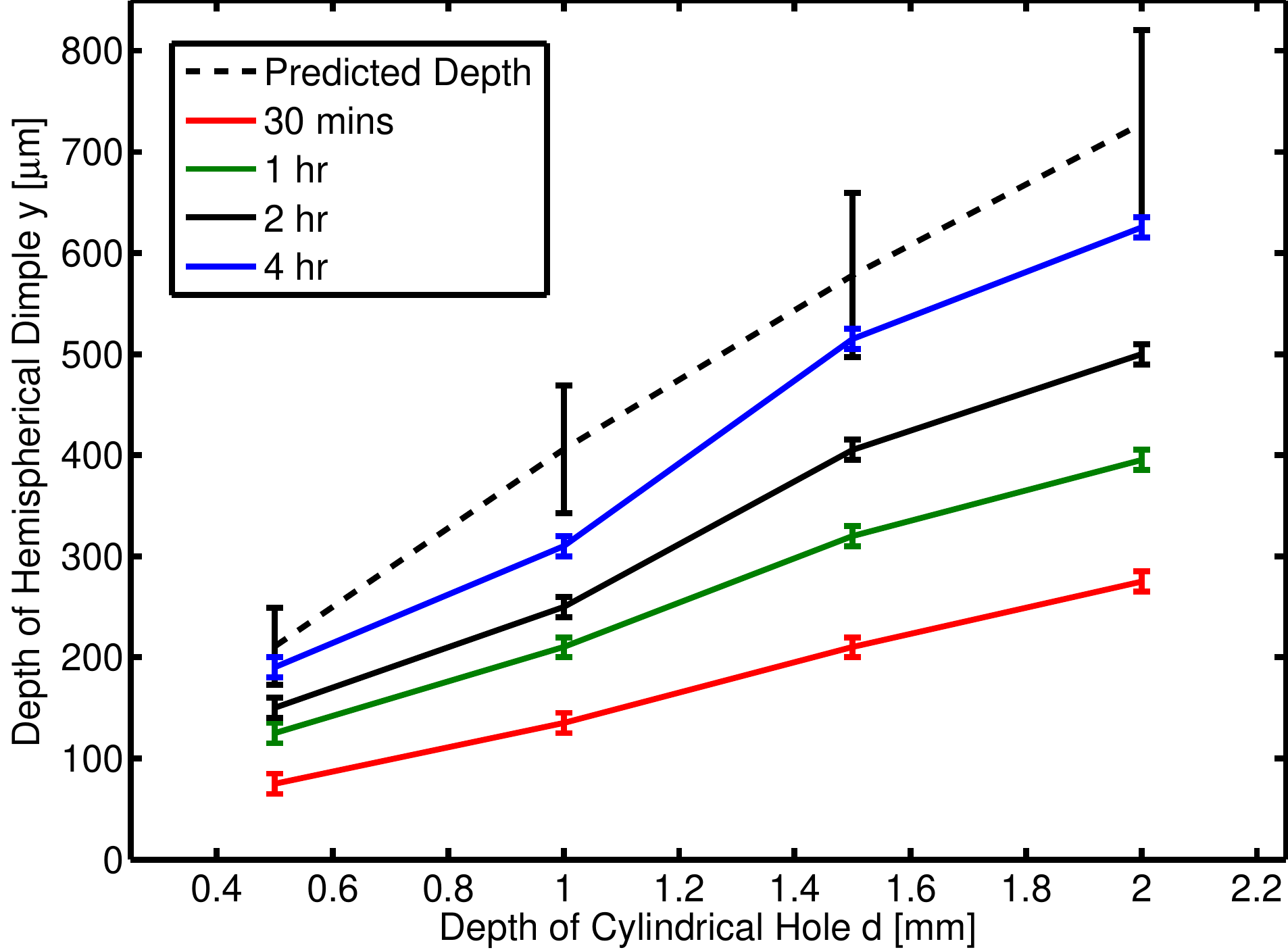}}
         \caption{\label{depthvshole} Depth of dimple $y$ measured for various depths $d$ of the cylindrical hole. The predicted depth is given by the dotted lines. The error bar on the predicted depth reflect the uncertainty in pressure. The fabrication time is varied ranging from 30 minutes to 4 hours.
    }
\end{figure}

For a fixed value of $r_0=1\,\mathrm{mm}$ and $\Delta P\sim 150\,\mathrm{mbar}$, we fabricate glass dimples for various hole depths and deformation times and we plot the depth $y$ in Fig.\ref{depthvshole}. Error bars on the theoretical plot reflect the uncertainty in pressure due to limitations of the furnace. Although the general trend of $y$ as a function $d$ is correct, there are two notable deviations from the model predictions: the measured depth does not converge to the predicted equilibrium value, and the time scale to reach equilibrium differs by more than an order of magnitude.

In the actual process, the sealing temperature is not known for sure. If it occurs below $800\,\mathrm{^oC}$, the enclosed gas in the cylindrical hole is heated up as the furnace reaches the final temperature and the pressure enclosed will be higher than we assumed. For example, if the hole is sealed at $700\,\mathrm{^oC}$ instead of the $800\,\mathrm{^oC}$, the pressure inside the hole will be larger than estimated by approximately $10\%$. This could account for the reduced equilibrium value of $y$.  It is not clear why there is a such a large discrepancy between the predicted rate of formation with the actual rate. However we do note that the viscosity has a very strong dependence on temperature.  In any case, the resulting deformation is very reproducible and the general trend allows us to estimate the parameters needed to achieve a particular mirror geometry.

\section{Mirror Characterization}

\begin{figure}[t]
\centerline{\includegraphics[width=1\textwidth]{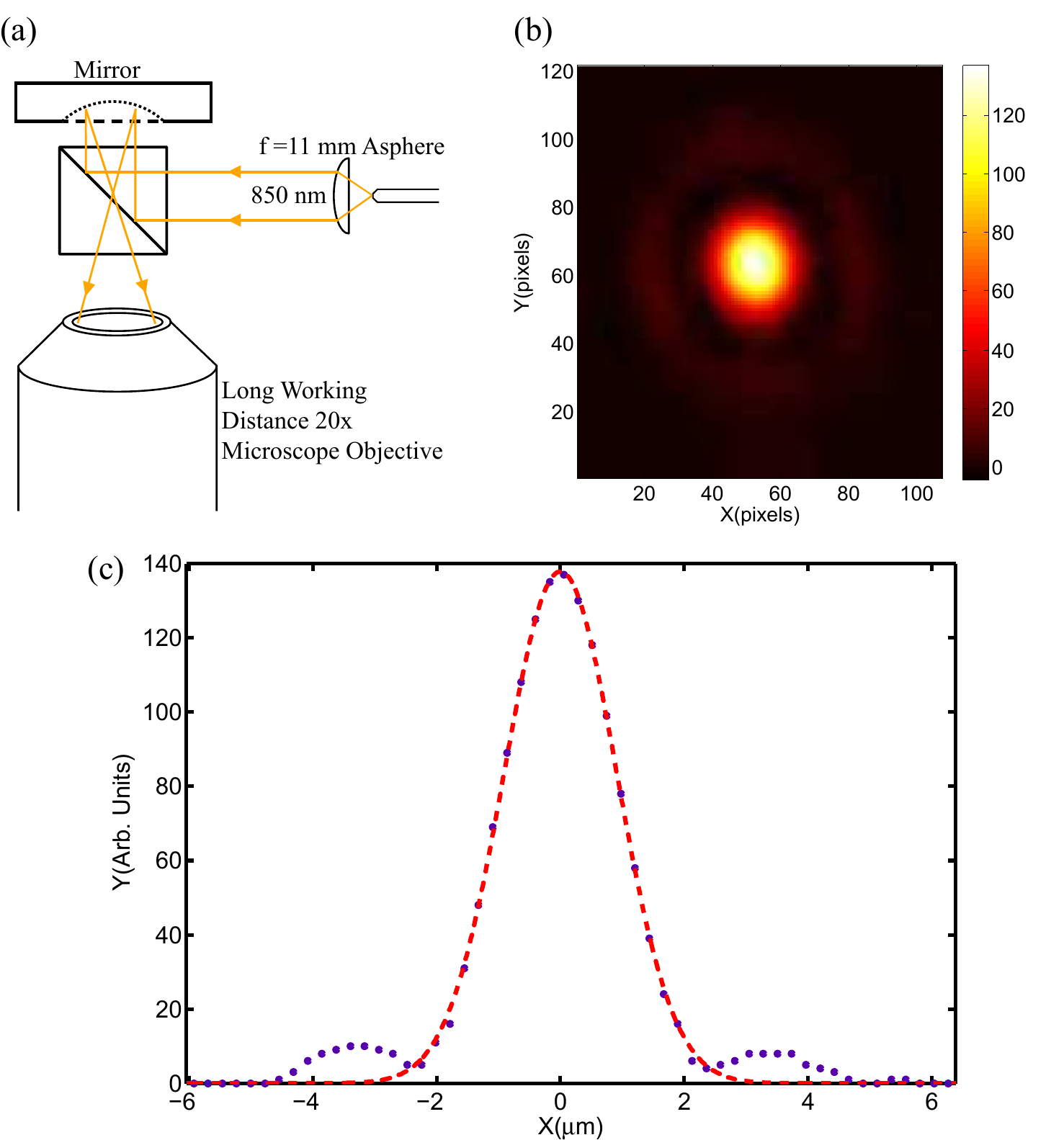}}
\caption{\label{spotprofile}(a) Setup used to measure the spot size of the reflected light from the spherical mirror. (b) Image of the spot showing the main peak and secondary rings. (c)  Cross section of the beam intensity showing a main Gaussian peak profile and smaller secondary ring.}
\end{figure}
There are just two independent parameters that determine the geometry of a spherical mirror, namely the clear aperture and the radius of curvature.  The clear aperture is given by the hole radius, $r_0$, and the radius of curvature, $R$, can be controlled by the pressure difference established during fabrication.  For our experiments we aimed for a clear aperture of $4\,\mathrm{mm}$ as this enables easy alignment of input beams with waists up to around $1\,\mathrm{mm}$ without significant clipping.  Allowing for a $1\,\mathrm{mm}$ working distance for the optic constrains the curvature to be about $3.3\,\mathrm{mm}$.  With this as a target, we fabricated a mirror which gave a measured curvature of $3.2\,\mathrm{mm}$ or, equivalently, a focal length of $1.6\,\mathrm{mm}$.  The mirror surface was then coated with evaporated silver giving a reflectivity greater than $95\%$ at both $780\,\mathrm{nm}$ and $850\,\mathrm{nm}$.

The focusing performance of the optic was tested using the setup shown in Fig.\ref{spotprofile}(a).  A collimated beam of waist $1\,\mathrm{mm}$ was incident on the mirror and the resulting focus was imaged using a calibrated 20x long working distance microscope objective. A 50:50 beam splitter was used as shown so that the focus could be imaged directly. Fig.\ref{spotprofile}(b) shows the actual focus imaged by the microscope objective. A strong central peak can be seen along with the presence of a secondary ring. Fitting the cross section of the image shown in Fig.\ref{spotprofile}(c) to a Gaussian profile gave a waist of 2 $\pm$0.1 microns. Subtracting a two dimensional Gaussian profile from the profile in Fig.\ref{spotprofile}(b), we then estimate, from the residuals, that 12\% of the total power lies in the outer ring.

Note that, although an ideal optic with a focal length of 1.6 mm would yield a sub micron spot size, spherical aberrations limits the achievable spot size and gives rise to the secondary ring. However, a spot size of 2 micron still enables us to use light assisted collisions to prepare a single atom \cite{grunzweig2010near}.

\section{Implementation of a tightly confining optical dipole trap}
In our experiment we implement a dipole force trap using light at $850\,\mathrm{nm}$.  The trap is loaded directly from a magneto-optical trap located at the focus of the spherical mirror and, after loading, we utilize light assisted collisions induced by the MOT cooling beams to obtain a single trapped atom.  In this section we discuss the details of our setup and the procedure for loading and detecting single atoms.

\begin{figure}[t]
\centerline{\includegraphics[width=0.7\textwidth]{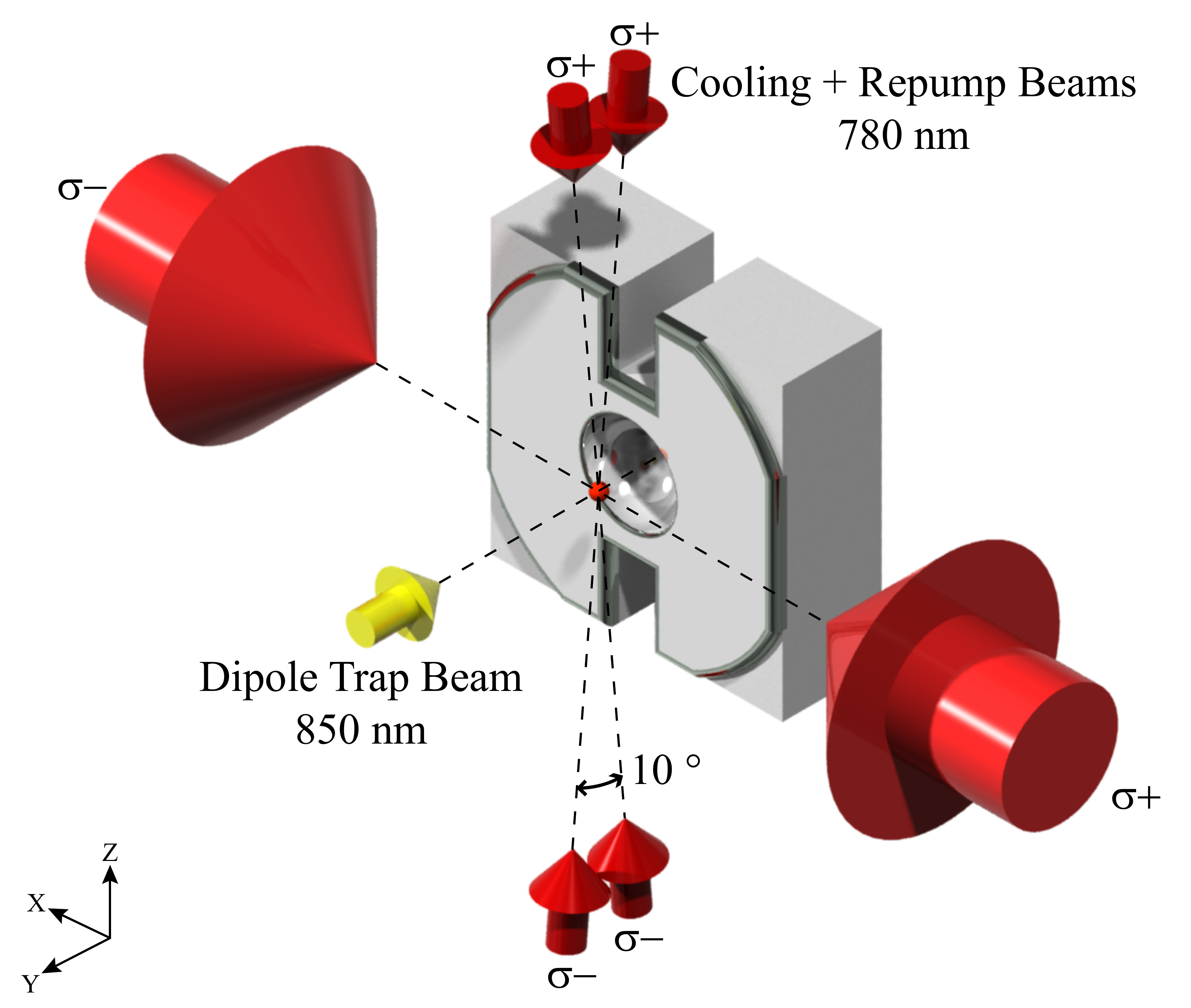}}
\caption{\label{expsetup} Experimental setup of the optical dipole trap. Red detuned 780 nm laser beams cool a cloud of atom at the focus of the spherical mirror. The mirror focuses 850 nm off-resonant light to a tight spot. The retroflected vertical cooling beams and the dipole trap beam lie on the YZ plane. The horizontal cooling beam is retroflected along the X direction.}
\end{figure}

\subsection{The Experimental Setup}
\begin{figure}[t]
\centerline{\includegraphics[width=0.8\textwidth]{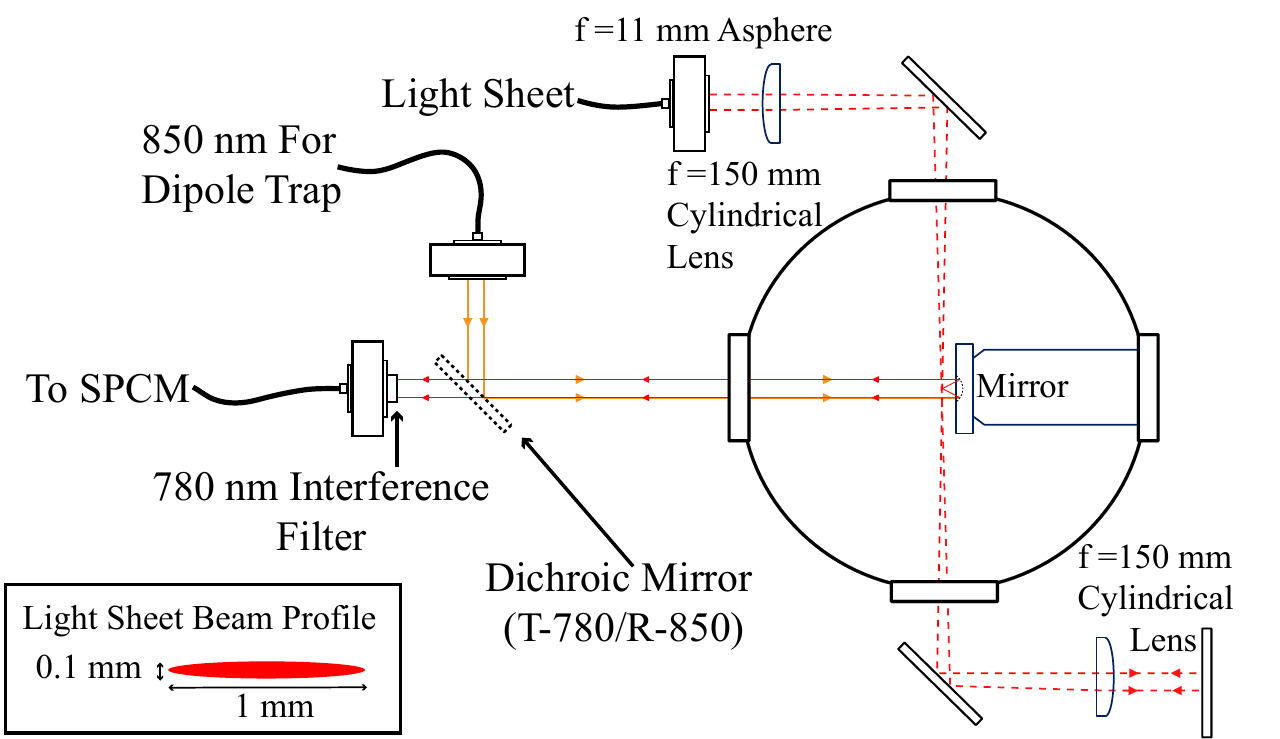}}
\caption{\label{optics} Optics setup for dipole trap and detection. Dipole trap light at 850 nm is combined along the collection beam path using a dual coated mirror. The dichoric mirror is reflective at 850 nm and transmissive at 780 nm. A retroflected light sheet is used to excite the atoms trapped at the focus of the dipole trap. The fluorescence light is reflected into a near collimated beam by the spherical mirror.(Inset)Beam profile of the light sheet.  A narrow band interference filter at 780 nm gets rid of spurious external light at the collection coupler.}
\end{figure}

Our experimental setup is illustrated in Fig.~\ref{expsetup} and Fig.~\ref{optics}.  The dipole force trap is formed using laser light at $850\,\mathrm{nm}$.  This light is obtained from an extended cavity diode laser which is spatially filtered by a single mode fibre. Light from the fibre is collimated with an aspheric lens giving a beam waist of $1\,\mathrm{mm}$ with a maximum available power of about $40\,\mathrm{mW}$.  The light is then combined along the path of the collection optics using a dichroic mirror as shown in Fig.~\ref{optics}.  Assuming a $2\,\mathrm{\mu m}$ focused waist as determined in the previous section, we estimate a trap depth of $2.5(2)\,\mathrm{mK}$ \cite{grimm2000optical}.

The trap is loaded directly from a magneto-optical trap located at the focus of the spherical mirror. We use a six beam MOT configuration consisting of two retroflected, near vertical beams and a third retroflected horizontal beam.  The near vertical beams have a waist of $500\,\mathrm{\mu m}$ with an angle of 5$^\circ$ from the vertical to facilitate making a MOT close to the surface of the optic.  The horizontal beam has a waist of $4.5\,\mathrm{mm}$ which gives maximum overlap with the intersection region of the two near vertical beams.  All cooling beams are red detuned below the (F=2 $\rightarrow$ F'=3), $D_{2}$ cooling transition. A re-pumping laser locked to the ($F=1 \rightarrow F'=2$) transition is mixed with the vertical beams. The magnetic field for the MOT is provided by circular coils in anti-Helmholtz configuration providing a magnetic field gradient of 20 G/cm. Under these conditions, we create a MOT with approximately 2 million atoms.

In our experiment we detect the presence of an atom using fluorescence measurements.  Since light from the cooling beams generates a large background contribution, we use an independent detection beam to excite the atoms. We use a linearly polarized light sheet which has a $1\,\mathrm{mm} \times 100 \,\mathrm{\mu m}$ cross-section. As illustrated in Fig~\ref{optics}, the light sheet is retro reflected to reduce the effects of radiation pressure during imaging. The power in the light sheet is such that it is at five times the saturation intensity of ($F=2 \rightarrow F'=3$) transition for isotropic polarization \cite{steck2001rubidium} and operated at a blue detuning of $55\,\mathrm{MHz}$.  This detuning compensates the AC Stark shifts induced by the dipole trap.

Light from the atoms is reflected into a nearly collimated beam by the spherical mirror. The light is focused using a 18 mm focal length ashperic lens into a multimode fibre connected to a single photon counting module. Based on the induced Stark shifts and the detuning of the detection light we estimate that the atom scatters approximately 700 photons during the $100\,\mathrm{\mu s}$ detection time out of which approximately 2\% are detected by the photon counting module. The 62.5 micron core diameter of the fibre also acts as a spatial filter cutting down on off-axis background light.

\subsection{Loading a single atom}
In general, the number of trapped atoms in the dipole trap, $N$ is governed by \cite{schlosser2002collisional}
\begin{equation}
\frac{\mathrm{d}N}{\mathrm{d}t}=R_{l}-\gamma N-\beta N(N-1),
\label{lightassisted}
\end{equation}
where $R_{l}$ is the loading rate, $\gamma$ is loss rate due to background collisions, and $\beta$ is the two body decay rate. Under typical vacuum conditions, the loss rate due to background collisions can be ignored and at equilibrium we have
\begin{equation}
R_{l}=\beta N(N-1).
\label{equilibrium}
\end{equation}
If $R_l\ll \beta$, the equilibrium number will be approximately zero or one atom. In such an experiment, one can monitor the trap continuously for the loading of a single atom and proceed with the subsequent experiment once a single atom is detected. However in our experiment, continuous monitoring of the dipole trap for a single atom is not possible as light from the MOT beams saturates our detector.  We use an alternate approach of using light assisted collisions with our cooling beams to get down to one atom and only one atom.

In the absence of loading, $R_{l}=0$, atom number decays according to the two body loss term. In presence of resonant light, two body losses rapidly eliminates atoms from the dipole trap until $N=0,\mbox{ or }1$.  At this point no further two body losses can occur.  To implement this, we first maximize our loading rate which typically gives $N\approx 15$.  We then switch off our magnetic coils, which reduces the loading rate to zero, and the cooling light is kept on for the next 100 ms to engage light assisted collisions. At the end of the $100\,\mathrm{ms}$, we check for the presence of an atom using a $100\,\mathrm{\mu s}$ pulse of light from our detection beam.

\begin{figure}[t]
\centerline{\includegraphics[width=1\textwidth]{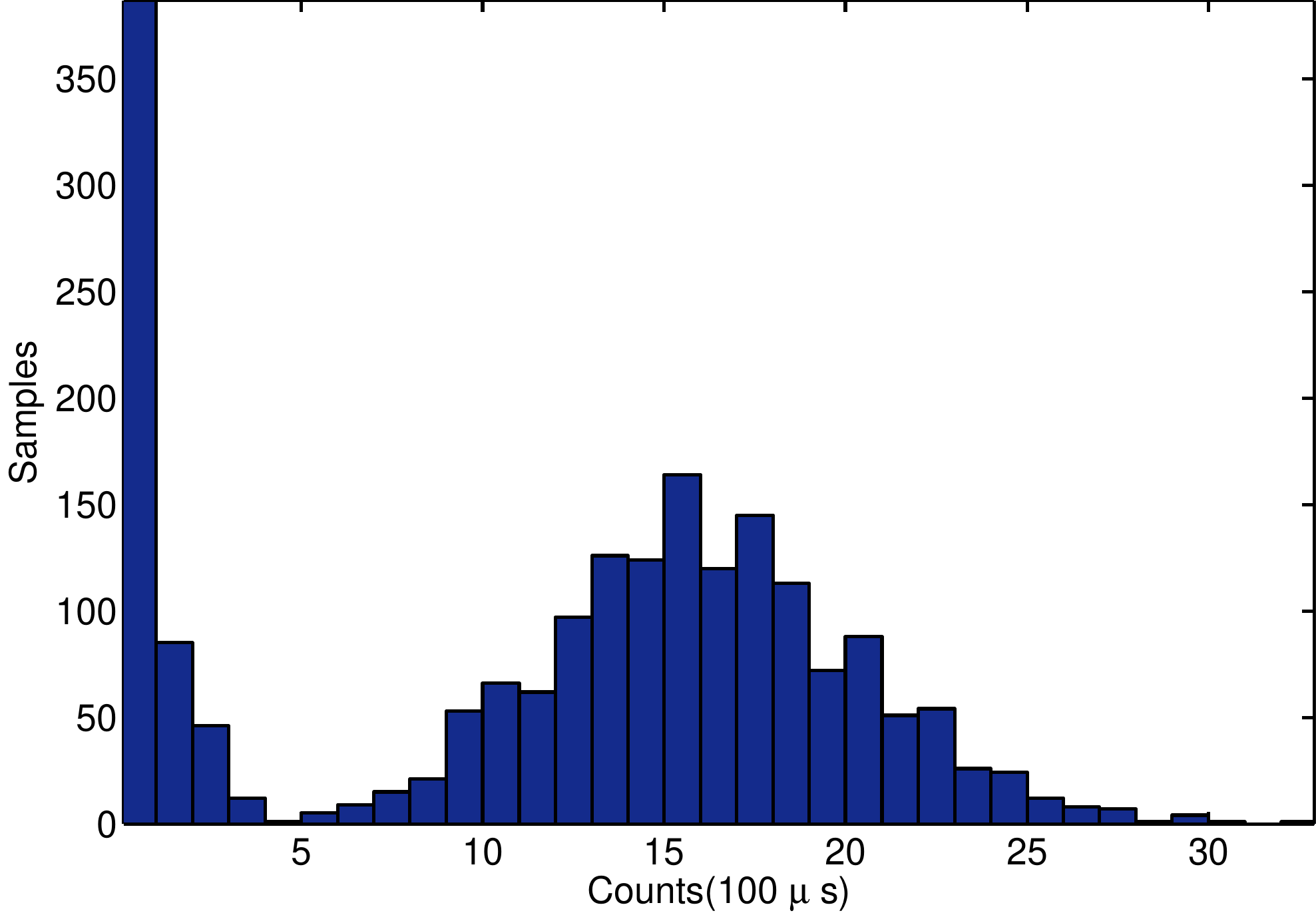}}
\caption{\label{singleatomsignal} Histogram of counts at the detector from a $100\,\mathrm{\mu s}$ excitation pulse with the light sheet. The histogram shows two Poisson distribution with means of approximately 16 counts and 1 count corresponding to single atom fluorescence counts and background counts.}
\end{figure}

The histogram of the counts detected from $2000$ experiments is shown in Fig.(\ref{singleatomsignal}).  It exhibits a bimodal Poissonian distribution with contributions from zero and one atom distributions. The distribution from zero atoms has a mean of 1 count per 100$\mu$s exposure time consistent with that obtained from independent background measurements.  The remaining part of the distribution is from a single atom and has a mean of 16 counts.  Based on this, we can detect the presence of an atom with greater than $99.7\%$ confidence and in greater than $70\%$ of the cases.

\section{Conclusion}
We have fabricated and demonstrated the use of a miniature spherical mirror to trap and detect a single atom. The mirror was easily integrated with a standard neutral atom experiment and has the dual function of tightly focusing an optical dipole trap as well as collecting light from the single atom. Using the system we have demonstrated a single atom detection efficiency of more than 99\% and single shot loading probability greater than 70 \% of the cases.

The simplicity of the setup along with the scalability of the fabrication process would make it suitable to create an array of addressable single atom traps. The accurate modelling of the process allows us to repeatedly fabricate mirrors to specific dimensions. Our experiment serves as a proof of concept for focusing a tight optical dipole trap using a spherical mirror. For future work, both the focusing and the light collection efficiency can be further improved by using a mirror of smaller dimensions.

\section{Acknowledgements}
We would like to thank Cecilia Muldoon of Oxford University for her suggestions which led to the success of this experiment and Markus Baden,Radu Cazan and Kyle Arnold for proofreading this paper. We would also like to thank Aarthi Dhanapaul for the mirror coatings and Joven Kwek for machining of the substrates. We acknowledge the support of this work by the National Research Foundation and the Ministry of Education
of Singapore, as well as by A-STAR under Project No.
SERC 052 123 0088.


\begin{thebibliography}{10}
\bibitem{Schott}
D263T is a highly resistant borosilicate glass manufactured by Schott.
\end{thebibliography}
\end{document}